\documentclass[sigconf]{acmart}
\usepackage{color,soul}
\usepackage{multirow}
\usepackage{array}
\usepackage{booktabs,multicol}
\usepackage{lipsum,graphicx,subcaption}

\AtBeginDocument{%
  \providecommand\BibTeX{{%
    \normalfont B\kern-0.5em{\scshape i\kern-0.25em b}\kern-0.8em\TeX}}}

\setcopyright{acmcopyright}
\copyrightyear{2018}
\acmYear{2018}
\acmDOI{10.1145/1122445.1122456}

\acmConference[Woodstock '18]{Woodstock '18: ACM Symposium on Neural
  Gaze Detection}{June 03--05, 2018}{Woodstock, NY}
\acmBooktitle{Woodstock '18: ACM Symposium on Neural Gaze Detection,
  June 03--05, 2018, Woodstock, NY}
\acmPrice{15.00}
\acmISBN{978-1-4503-XXXX-X/18/06}

\begin{document}

\title{COVID Induced Digital Inequality for Senior Citizens}


\author{Nicky Qiu}
\email{nickyqiumail@gmail.com}


\begin{abstract}
The global pandemic of COVID-19 has fundamentally changed how people interact, especially with the introduction of technology-based measures that aim at curbing the spread of the virus. As the country that currently implements one of the tightest technology-based COVID prevention policy, China has protected its citizen with a prolonged peaceful time of zero case as well as a fast reaction to potential upsurging of the disease. However, such mobile-based technology does come with sacrifices, especially for senior citizens who find themselves difficult to adapt to modern technologies. In this study, we demonstrated the fact that most senior citizens find it difficult to use the health code apps called ``JKM'', to which they responded by cutting down on travel and reducing local commuting to locations where the verification of JKM is needed. Such compromise has physical and mental consequences and leads to inequalities in infrastructure, social isolation and self-sufficiency. As we illustrated in the paper, such decrease in life quality of senior citizens can be greatly reduced if improvements on the user interactions of the JKM can be implemented. To the best of our knowledge, we are the first systemic study of digital inequality due to mobile-based COVID prevention technologies for senior citizens in China. As similar technologies become widely adopted around the world, we wish to shed light on how  widened digital inequality increasingly affects the life quality of senior citizens in the pandemic era.

\end{abstract}



\maketitle

\section{Introduction}
In early 2020, COVID-19 hit the world by surprise, and has quickly evolved into a global pandemic. Nowadays, although vaccines have been widely available, the unenforceable national vaccination rates and the possibility of vaccine penetration makes disease prevention a long term mission for at least the foreseeable future. As COVID ranks as one of the most transmissible disease that mankind has ever discovered \cite{wang2021covid}, a key measurements to prevent spreading is contact tracing: locating people who have had close contact with those infected. Mobile-based technology, with its wide availability and location tracking capability, is thus chosen to estimate each individual's risk of exposure to the disease. As a result, mobile tracking tools are developed and applied in many countries around the world, including the United States, the United Kingdom\cite{wymant2021epidemiological}, China \cite{liang2020covid19, zhou2021lessons, chen2021health}, and Saudi Arabia \cite{hidayat2021mobile}.

Suffered from the first and one of the deadest pandemic, China has adopted a zero-tolerance disease control with a mobile app-based health codes checking system: a health code called ``JKM'' that evaluates the risk of an individual based on the testing history and contact history to those infected, and a ``trajectory code'' that locates the individual and tracks their travel history. Each province has its own JKM that incorporates the same guidelines from the central disease control system in addition to other risk factors determined by the local authorities. Each individual is required to show his/her JKM and sometimes in combination of trajectory code in order to enter almost all public area throughout China. Such mobile based technology has been proven effective as China has many of the recent outbreaks under control.

The widely successful digital-based COVID policy implemented in China, however, does not come without sacrifice. People who do not use smartphones are unable to realize location tracking or JKM display, thus unable to enter most public places. This raises not only physical, but also psychological concerns in an already stressful event like COVID. In addition, JKM is developed and integrated into two of the most widely used Chinese mobile applications: Wechat and Alipay. These two mobile apps, however, are designed for social networking and mobile payment, respectively, and do not have a dedicated user interface for JKM. Such design makes it even harder for senior citizens who, on average, are already not as good at using smartphones. This will aggravate the already widened digital inequality for senior citizens in China.

We should not neglect the obstacles that seniors encounter in everyday life and in travel when using the mandatory JKM, as major countries continue to experience population aging. In 2020, China had 190.6
million adults aged 65 and over, accounting for 13.5 percent of the total population \cite{jin2022association}, and this number continues to increase. More usable JKM will benefit a large portion of the population on a daily basis and reduce digital inequality. 


In this paper, we analyzed the extend to which senior citizens' everyday lives are affected by the JKM app in China. Our results showed that both their travel intentions and their willingness to commute to local activity centers and grocery stores are significantly reduced. This illustrated senior citizens' adaptive behavior to the complexity of the interactive process of the JKM. Further analysis indicated that such phenomenon had consequently reduced the sense of social acceptance, confidence, and self-sufficiency of senior citizens in China, making them more and more reliable on their children, who can usually operate the JKM more easily. Moreover, as the trajectory code records a person's travel history and as senior citizen usually relies on others to show their JKM, privacy concerns arise which further prevents them from traveling.  

Fortunately, our analysis showed that most frustrations of the senior citizens on using the JKM app can be alleviated by a better user experience design. For example, our survey indicated that a dedicated JKM app that does not reside in any of the exiting apps would significantly improve the user experience. Others believe that user interface improvements such as large fonts and better layout will also help. In addition, alternative ways to acquire the health verification such as phone calls and sms will also significantly improve the experience of senior citizens.
 
\section{Backgrounds}
We review some of the background to this work, ranging from digital inequality induced from the adaptation of the technology driven society to the potential impacts that COVID have brought to for senior citizens. We also review a list of mobile based contact tracing applications across the world.  
\subsection{Digital Inequality}
Technology especially those aimed at facilitate communication such as Internet and mobile devices, has made it significantly easier for people to connect and the interact. As the society transforms to be more and more relied on those technology, people who are more adapted to take advantage of these tools became more privileged in this setting over those who are not, creating what we called digital inequality\cite{dimaggio2001digital,robinson2009taste,dimaggio2004unequal,kvasny2006cultural}. Seniors \cite{friemel2016digital}, low income families\cite{hsieh2008understanding}, and minorities \cite{d2013internet, katz2017digital} are more likely to be affected as their access to the technology and the way how they use it is often limited compare to other groups \cite{dimaggio2004digital}. As the majority of the major countries entering a decline in fertility rate and consequently a rapid growth of the aging population, digital divide for the aged is becoming a pressing societal issue. Older adults are highly heterogeneous in terms of technology use \cite{van2017diversity}, which are mostly impacted by their prior employment experiences. Research studies have showed that old adults tend to develop their own routines in using technology \cite{quan2016interviews}, which suggested the need of interactive designs specifically targeted at old adults. 

\subsection{Pandemic Impacts on Technology Use}
The COVID-19 pandemic have fundamentally shifted how people use technology. As in-person communications being significantly reduced, either as a personal precaution or as a government enforced disease prevention policy, more and more interactive experiences are moved from offline to online. The use of technology will significantly affect the quality of the interactions ranging from online learning \cite{adnan2020online, chakraborty2021opinion, liguori2020offline}, information seeking \cite{reisdorf2021information}, health care \cite{okonkwo2021covid}, work \cite{alaqra2021impact}, to communications \cite{alaqra2021impact}. As we get more isolated compared to prior pandemic, people seek help from online to deal with COVID-19 Loneliness \cite{niu2021stayhome}. Privacy concerns raise as more interactions are exposed to the digital tools than ever \cite{niu2021stayhome, seberger2021us}. And the pandemic seem to impacts the old adults more than other age groups \cite{van2021loneliness}. Needless to say, the unique challenges of using technologies in the pandemic era call for changes in designing technology interface for better access. 

\subsection{Mobile Based Contact Tracing}
A key approach in curbing the COVID pandemic is the quarantine of the infected individuals along their close contacts. Smart phone devices, due to their wide availability and their GPS tracing capabilities, become an ideal tool to conduct contact tracing. Various App based contact tracing tools are developed \cite{kondylakis2020covid, adeniyi2020mobile}. Many of them are developed officially by the local authorities include the JKM System by the Chinese Government \cite{liang2020covid19, zhou2021lessons, chen2021health}, Tawakkalna by Saudi Arabia \cite{hidayat2021mobile}, NHS by the UK government \cite{wymant2021epidemiological}, and COVIDSafe by Australia \cite{yang2021comparative}. As with any other technologies, old adults often found themselves difficult to adopt to it compared to other age groups. In countries (e.g, China) where mobile based apps are required to access most of the public spaces, the access to those places from senior citizens are also greatly constrained. As a results, many research calls for studies of impacts of such technology to old adults \cite{wang2021impact} and many early results have find preliminary evidences  \cite{alharbi2021pandemic, zhu2021user} on the pain points seniors experience when interacting with these technologies. 

\section{JKM System Revisited}
\begin{figure}[h]
    \centering
    \includegraphics[width=0.5\textwidth]{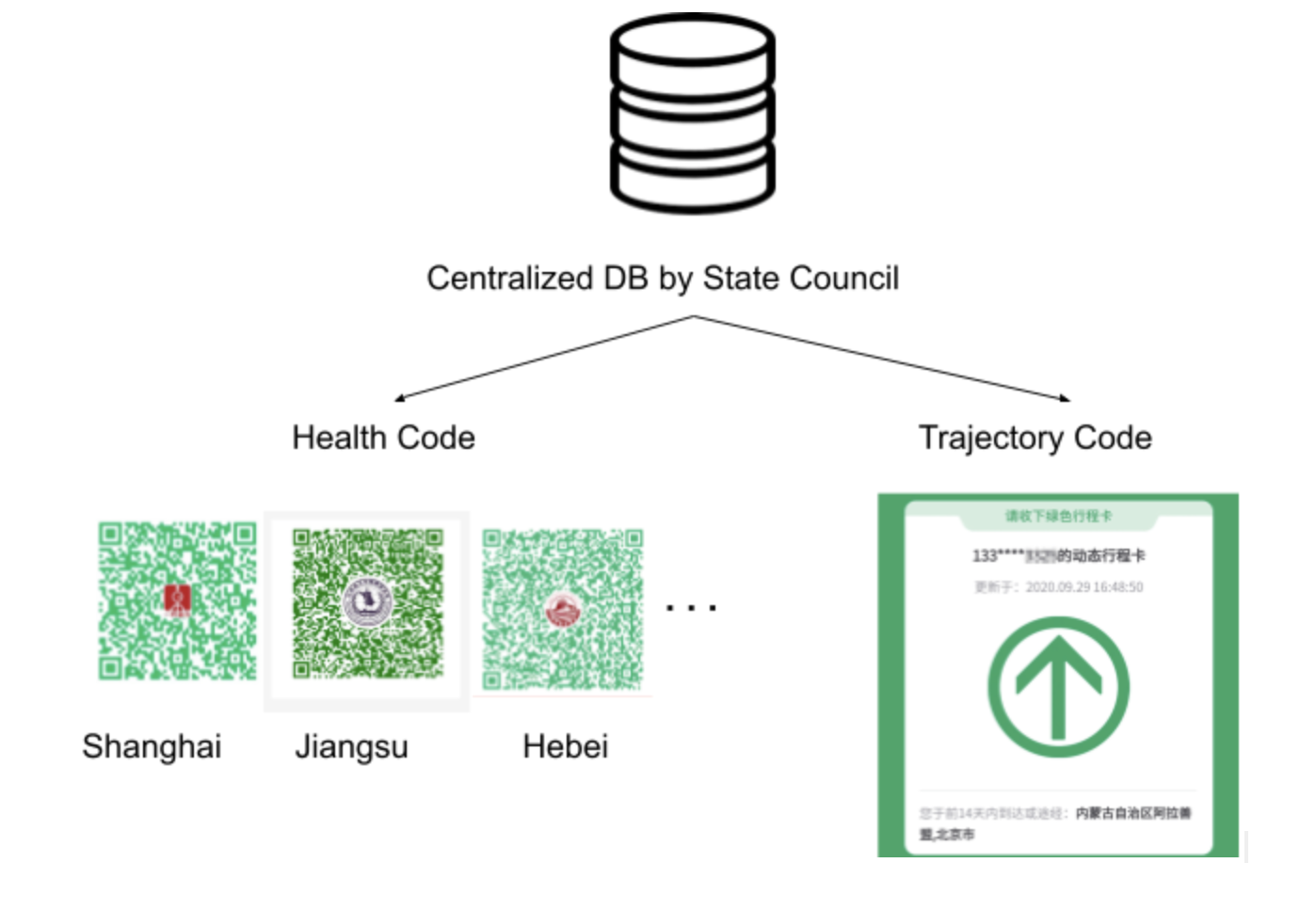}
    \caption{Example of the local JKM (left) and the trajectory code (right).}
    \label{plot:pic1}
\end{figure}

The Chinese JKM is a complex system consists of a centralized database and two set of QR codes: JKM and Trajectory Code, as illustrated in Fig.~\ref{plot:pic1}. Those two QR codes are used in combination to determine admission to public facilities as well as the need for quarantine.

\subsection{JKM}

\begin{figure}[h]
    \centering
    \includegraphics[width=0.5\textwidth]{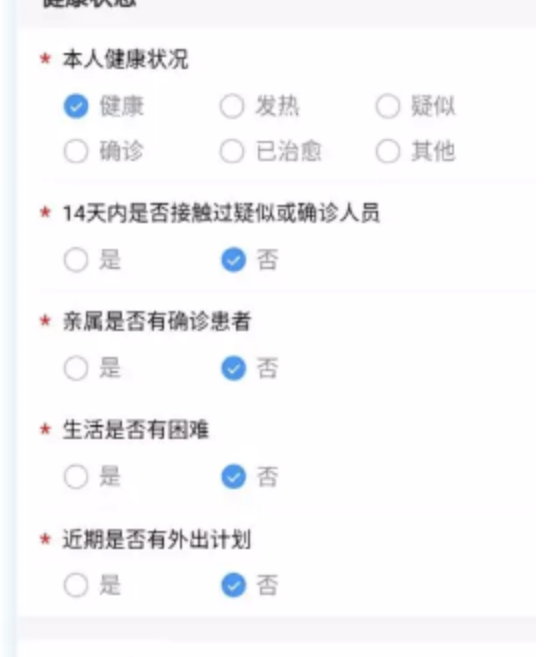}
    \caption{Questionnaire for Acquiring JKM for a Different Province }
    \label{plot:pic2}
\end{figure}

\begin{table*}
\begin{center}
\caption{Major Survey Questions}
\begin{tabular}{  l | c } 
\toprule
\textbf{Question} & \textbf{Answer} \\
\midrule
* Are you satisfied with JKM’s design? & Rate on scale of 1 to 5\\
* How to make JKM easier to use? & A. Enlarger text; B. Make JKM an individual APP;\\
& C. Text or call $\#12345$ to receive JKM status; D. Other\\
* If COVID risk rating is low, what is your willingness to go & \\ 
\hspace{0.25cm} to essential locations (supermarket/bus/hospitals…)?& \\
\hspace{1cm} When JKM is not checked & Rate on scale of 1 to 5\\
\hspace{1cm} When JKM is checked & Rate on scale of 1 to 5\\
\hspace{1cm} When there is a more convenience way of checking JKM, &\\
\hspace{1cm} or someone helps you find your JKM & Rate on scale of 1 to 5\\
*If COVID risk rating is low, what is your willingness to go to& \\
\hspace{0.25cm}leisure locations (friends places/malls/dances/extracurriculums…)?&\\
\hspace{1cm}When JKM is not checked & Rate on scale of 1 to 5\\
\hspace{1cm}When JKM is checked & Rate on scale of 1 to 5\\
\hspace{1cm}When there is a more convenience way of checking JKM,&\\
\hspace{1cm} or someone helps you find your JKM & Rate on scale of 1 to 5\\
*If COVID risk rating is low, what is your willingness &\\
\hspace{0.25cm} to travel to other cities?&\\
\hspace{1cm}When JKM is not checked & Rate on scale of 1 to 5\\
\hspace{1cm}When JKM is checked & Rate on scale of 1 to 5\\
\hspace{1cm}When there is a more convenience way of checking JKM, &\\
\hspace{1cm} or someone helps you find your JKM & Rate on scale of 1 to 5 \\
*Please rate your proficiency in phone operation & Rate on scale of 1 to 5\\
*What are the 3 APP’s that you use the most? & [open ended] \\
\bottomrule
\end{tabular}
\label{tab:survey}
\end{center}
\end{table*}

The JKM provides a single entry to illustrate an individual's health condition related to COVID based on information in the central government's database. There are three status of the JKM: Green, yellow, and red, indicating different level of risks for that specific individual. Information will be used when JKM is determined includes: 1) COVID testing results (if any), 2) vaccine status, and 3) whether a person is considered close contact to an infected individual based on data from trajectory code (discussed in the next section), and 4)other situations specified by the local or central government (e.g., when mandatory COVID testing is enforced JKM from anyone who did not conduct the test will be overridden to be red and will not resume until the results are negative). One unique characteristic for the Chinese JKM is that each province will have its own standard in determining the status of the JKM based on the same set of input data, as illustrated in Fig.\ref{plot:pic1}. This mechanics provides flexibility for the provinces to set their own disease prevention rules. When traveling to a different province, a questionnaire will have to be filled for a set of additional questions in order to register for the JKM belongs to a different states. An example has been provided in Fig.~\ref{plot:pic2}. Many of the questions are multiple choice ones asking the exact travel destination as well as the health condition and the the contact history within the past 14 days. One will have to fill out the questionnaire each time when he/she visits a new province regardless of whether that province has been visited before. In our analysis in the later part of the paper, the registration procedure set a major obstacle to senior citizens as fonts in the phones are usually too small for most to recognized and the survey is usually filed in the boarder of the new province where lots of travellers waiting in line. 

\subsection{Trajectory Code}
\begin{figure}[h]
    \centering
    \includegraphics[width=0.5\textwidth]{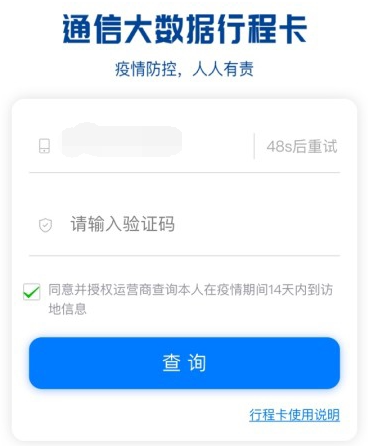}
    \caption{Verification Process for Trajectory Code}
    \label{plot:pic3}
\end{figure}

The trajectory has dual functions: 1) it exists as a data collection mechanism to acquire user's geo-spatial locations, and 2) it can be used to illustrate an individual's health status similar to the JKM. As the GPS location can often be forged by providing software simulated locations, the Chinese trajectory code uses cellular signal to locate an individual's location. In order to acquire the trajectory code, one will have to enter his/her Cheese mobile phone number in order to receive an verification pin code, which interface is illustrated in Fig.~\ref{plot:pic3}. The geo-location will be inferred using cellular services which is usually much different to forge. Once the trajectory code is requested, one will have to verify it through sms messages, in which case the system will collect a data point of presence and will be used later in constructing the actual travel trajectory. Such a practice might be very common for young adults who are constantly using online services that require phone verification. However, for senior citizens this procedure can be difficult to achieve as they will have to fill in the phone number, click on ``authorizing to release the information'' checkbox and request and fill in their verification code. In our interview, we find many senior citizens difficult to locate and the understand the content on the page due to the relatively small font size and the fact that most old adults have presbyopia, which prevented them from viewing close items clearly. 
\subsection{Software Entry Points}
JKM can be accessed through either Wechat, Alipay, or a service app provided from the government of each province. JKM is developed as a service or a function within each of the platform, and the entry point of each platform varies. For wechat, which is the most widely used apps for accessing JKM in our experience, is a social networking apps. In order to access JKM, one will have to either scan the QR code or to search for the name under the in-app developed program section. In Alipay, which is a payment app, one can access to the JKM in a similar way but the search procedure is different. In the case of the official apps provided by the government, it is developed as a service of that app and access methods vary for each province. As we will later show in our analysis, the lack of independent app is one of the main complains from senior citizens. Improving the interactive experience of the users will significantly improve senior citizens' will to travel and to perform their day-to-day job. 

\subsection{Facility Admission and Mandatory Quarantine}
JKM is used to determine the access to public facilities and the risk of mandatory quarantine in China. Only the green code will guarantee access to public facilities. This means that JKM will be checked upon every visit to public facilities such as building, markets, transportation, hospitals, schools, and cinemas. JKM is usually the primary JKM to check. In case where the local disease becomes a bit severe, trajectory code will also needed to be checked. Mandatory quarantine will be conducted based on the local authority's assessments of individual risks, usually when the travel trajectory insects with those who have been infected.

\section{Methods}
We conducted surveys and interviews among senior citizens in Shanghai to understand the impact JKM has made on the aspects of their lives. 

\subsection{Data Collection}
We recruited participants from 5 different locations in Shanghai, China: Shanghai Old-Age University, Xiangyang Park, Jiangpu Park, Taichi class, and Golden Harbor Apartment. We picked these locations because they are spread out across Shanghai, and they are where seniors with various backgrounds come and gather.

\subsubsection{Survey}.



We designed the survey to make it accessible to our target audience in order to minimize the systematic bias. The survey is made in paper format and distributed in person, therefore seniors who are unable to fill out online surveys are not discouraged or excluded from the population. The fonts and color contrast are increase so as to make it more readable for vision degenerated seniors. For seniors who are still not able to read due to vision or illiteracy, the survey is read to them. To hear the voice of seniors with various academic backgrounds, we phrased the questions in everyday terms and minimized barriers of academic ability such as ability to read charts. The survey takes about 15 to 20 minutes to complete. We demonstrated some of the major survey questions in Tab~\cite{tab:survey}.

\subsubsection{Interview}.

We conducted interviews in the same 5 locations as survey with 11 seniors who are willing to share their opinions. We observed organic conversations among participants and encouraged participants to liberally think out loud. Instead of rigorously structured Q \& A, we only ask questions occasionally to facilitate the discussion and have our key questions answered. The answers are not recorded due to participants’ concerns of privacy, but are transcribed for research purposes. The interviews last from 20 minutes to 50 minutes. We have summarized the list of interview questions in Tab.~\ref{tab:interview}.


\begin{table*}
\begin{center}
\caption{Interview Questions We Asked On-site.}
\begin{tabular}{ l } 
 \toprule
*What kind of JKM do you use?\\ 
 \hspace{0.5cm}(if participant does not use JKM) \\
 \hspace{0.5cm} how do you go to places which checks JKM, such as the hospital?\\ 
*What is the impact of JKM on your willingness to travel?\\ 
*How often do you use your cell phone?\\
*What APPs do you use most often?\\
 \bottomrule
\end{tabular}
\label{tab:interview}
\end{center}
\end{table*}

\subsection{Dataset Overview}


\begin{table*}[!htbp]
\centering
\caption{Demographics Statistics of Our Survey Data}
\begin{tabular}{c c|c c|c c|c c|c c}
\toprule
\multicolumn{2}{c}{Gender} & \multicolumn{2}{c}{Age}&
\multicolumn{2}{c}{Education}&
\multicolumn{2}{c}{Platform}&
\multicolumn{2}{c}{Live with?}\\
\midrule
Female & 62\% & 55 - 59 & 23\% &
      Under Primary School & 8\% & 
      IOS   & 54\% & Self&8\% \\
Male & 38\% &  60 - 69 & 46\% &
      Primary School & 8\% & 
      Android  & 15\% & Partner & 62\% \\
     &      & 70 - 79 & 8\% &   
      Middle School & 23\% & 
      Others  & 31\% & Children&23\% \\
    &       & 80 - 89 & 23\%  &
     High School & 54\% &
    &        & Grand children and children	&8\% \\
\bottomrule
\end{tabular}
\label{tab:demographics}
\end{table*}

Our sample was diverse in terms of demographics (see Table~\ref{tab:demographics} for key demographics). The 8 female and 5 male participants ranged in age from 58 to 83, living by themselves, with partners, children, or grand children. They have educational backgrounds ranging from under primary school to college, and occupations including accountants, government officials, professors, workers, servers and so forth. About three quarters of them use a smartphone, such as Apple, Huawei or others. At the time of our survey, two-thirds of our participants uses JKM through an APP, and others find an alternative way when JKM is checked.

\section{Shifts in Travel Patterns by JKM}
Senior citizens' travel patterns changed significantly in the era of COVID. Part of this is inevitable as seniors are at greater risks of developing severe symptoms of COVID compared to other age groups. Another part of the story has to do with the design of the JKM, which created an extra barrier for seniors to travel. In this section, we analyze the results from survey and interview data and study how the JKM itself other than the COVID affects the travel patterns of the Seniors.

\subsection{Survey Results}

\begin{figure*}[t]
\begin{multicols}{3}
    \includegraphics[width=.33\textwidth]{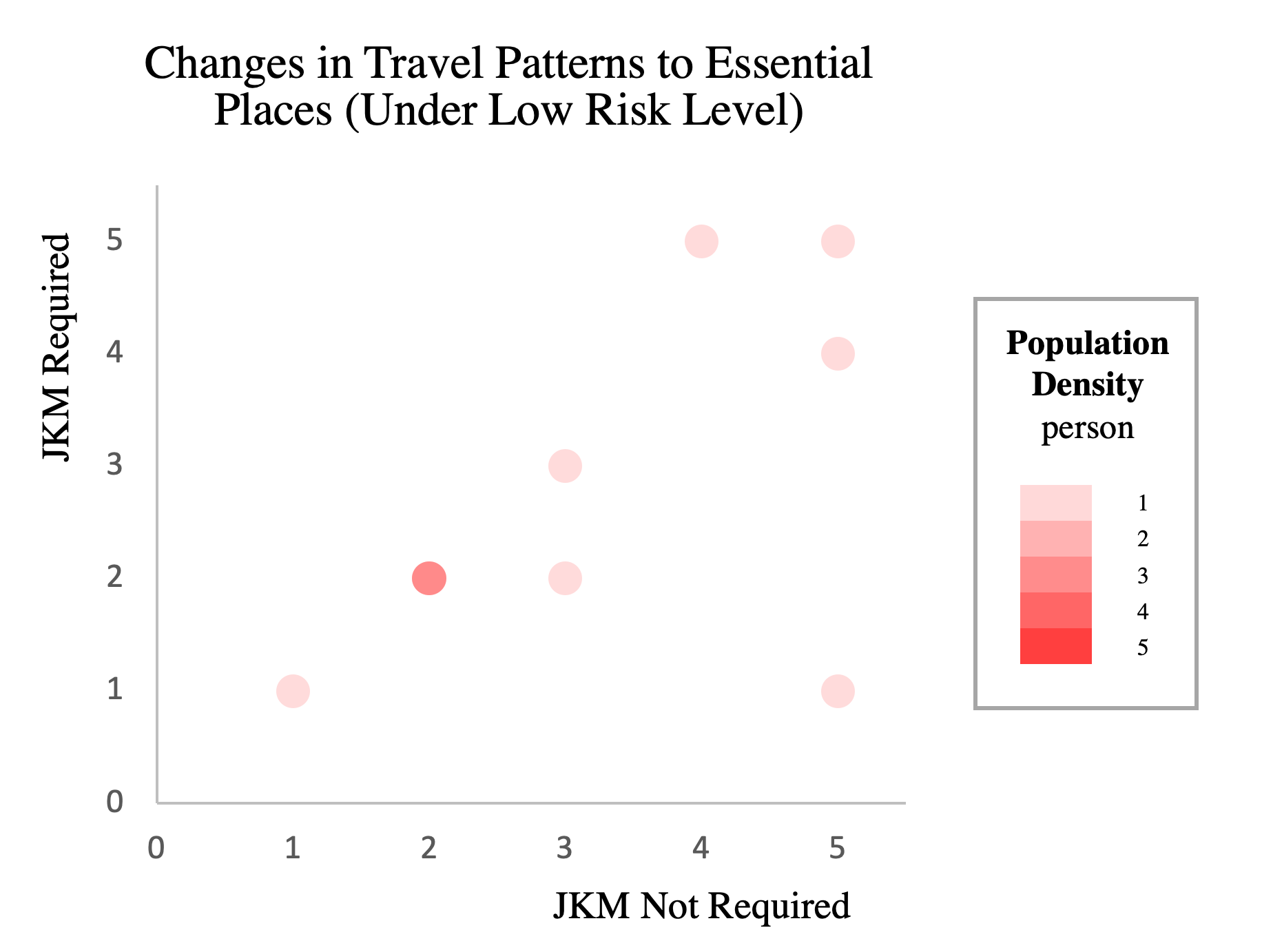}\par
    \includegraphics[width=.33\textwidth]{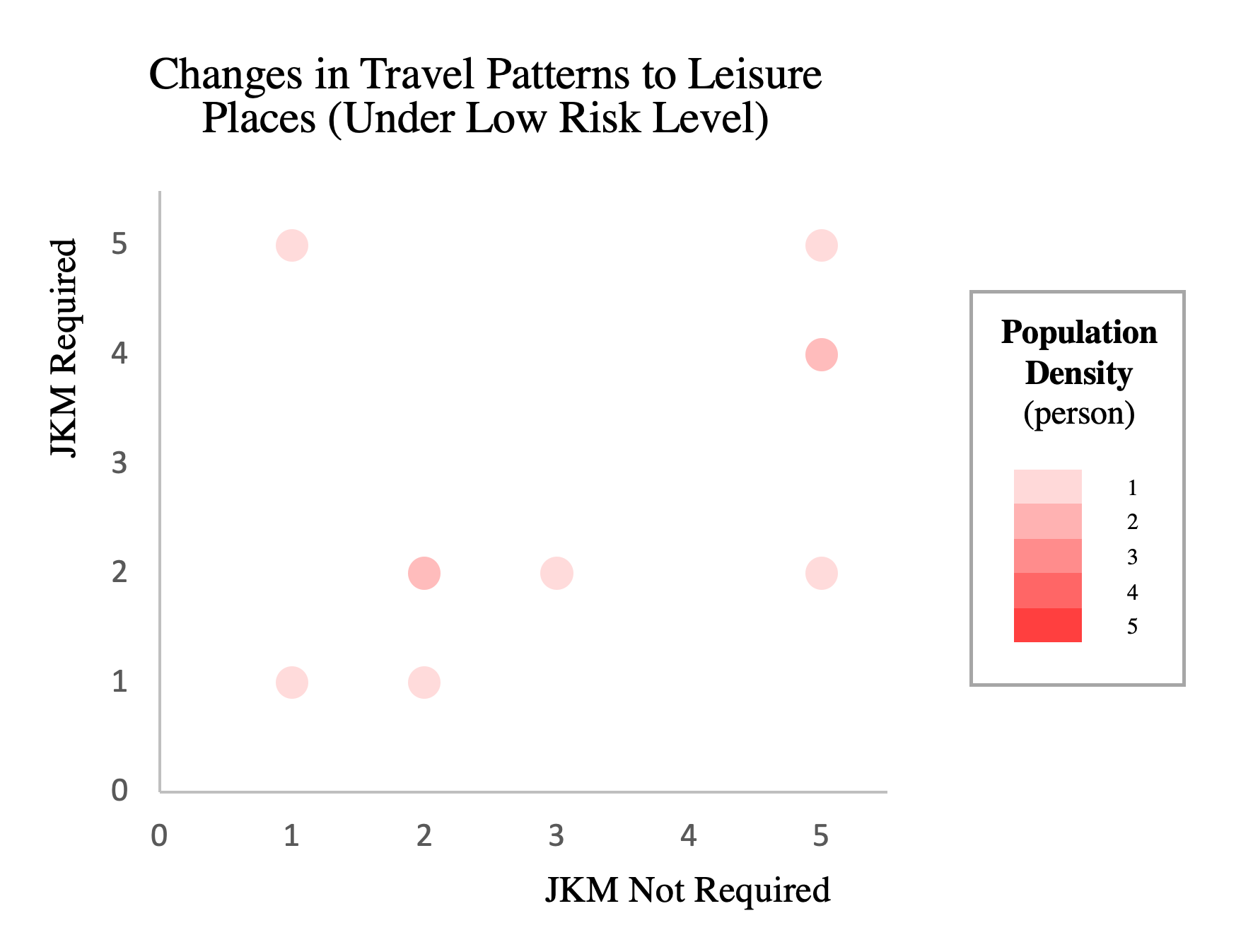}\par
    \includegraphics[width=.33\textwidth]{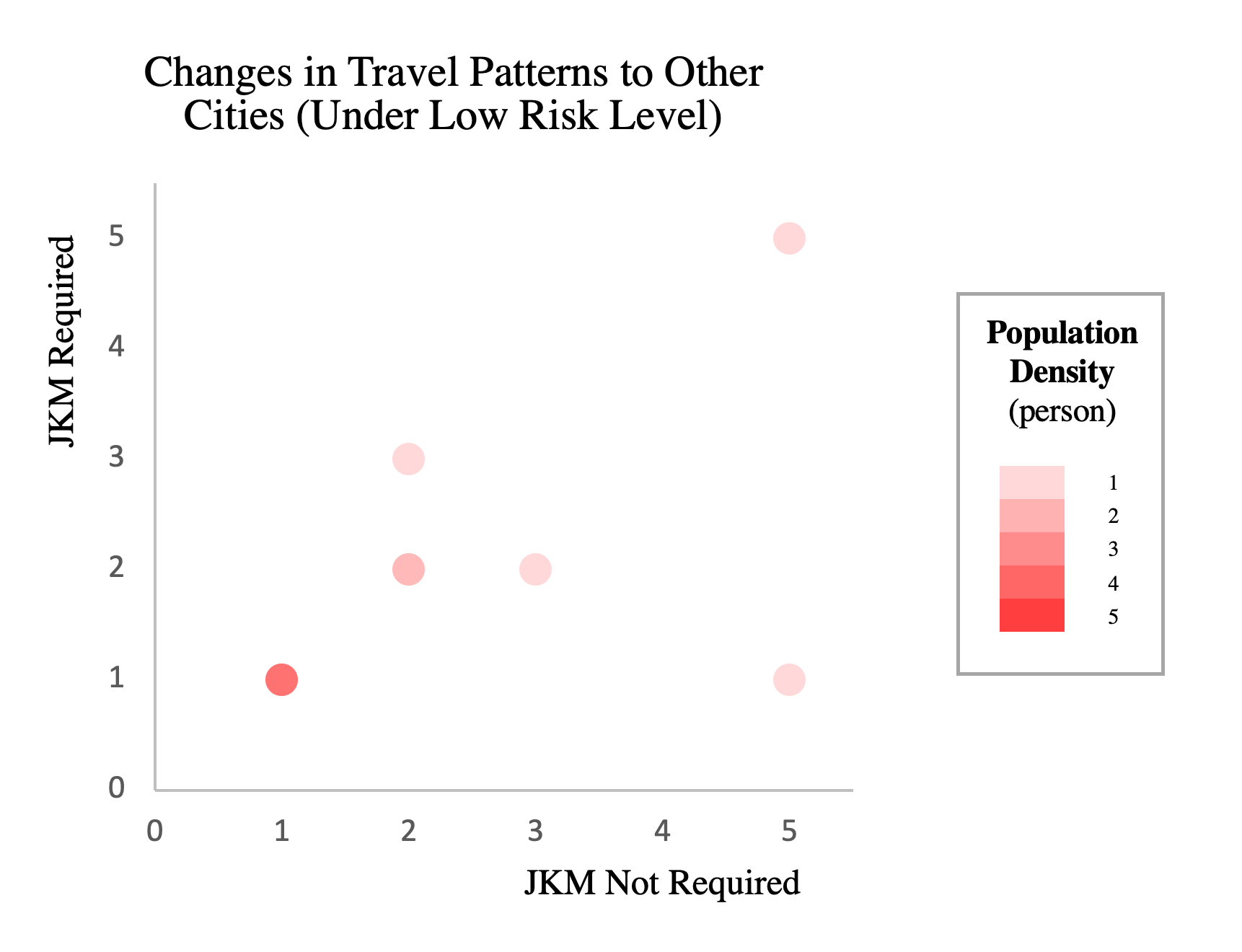}\par
\end{multicols}
\caption{Changes in Travel Patterns with JKM Required  and JKM not Required, Under Low Risk Level, to a) Essential Places, b) Leisure Places, and c) Other Cities}
\label{plot:ncvc1}
\end{figure*}

We analyzed the results from the survey on how much JKM changed senior citizens' willingness of traveling. This survey assumes that all travel destination is classified as ``low risk'' \footnote{defined as an area that has accumulative COVID cases of 50 or below in the last 14 days. } by the Chinese government, as indicated in our survey questions in Tab~\ref{tab:survey}. This assumption of made sure that the changes in travel intentions are not caused by COVID itself. We categories the destinations into three distinct types: a) essential places (e.g. supermarkets, hospitals, subway stations etc), b) entertainment (e.g. friend’s places, malls, parks, extracurricular classes etc) and c) tourism, which destination usually lies in another city. 

We compared the data with/without checking JKM, and the results are illustrated in Fig.~\ref{plot:ncvc1}. Here we see that travel intentions for each category falls into 0-5, and a reference line can be drawn with $x=y$ if no travel pattern changes can be recognized. Any points above $x=y$ means that the checking JKM makes senior citizens more willing to travel to a destination. On the other hand, data points below $x=y$ means that checking JKM creates extra hurdles to prevent old adults from going to these locations. Overall, participants’ desire to travel decreases when JKM is checked versus not checked, with everything else held equal. Out of the three hypothetical situation, the desire to travel to essential places such as supermarkets and hospitals are least impacted. On the one hand, like many participants said “you have to go, you have to go”. On the other hand, seniors still choose to decrease their frequency of travel when JKM is checked. Fig.~\ref{plot:ncvc1} also shows that places for entertainment activities, such as friend’s places, malls, parks, extracurricular classes etc are most impacted. This is because participants consider these places as “do not have to go”, thus would choose to go less often when these places are less accessible. Although it does not hurt their basic living and health needs the way essential locations do, it does contribute to their social isolation and mental well being in ways that are less noticeable. 

\subsection{Reduced Access to Healthcare}
In our interview, we also evaluated the degree to which the respondents' access to hospitals and healthcare services are affected. One participant said she does go to the hospital “as long as I could ensure [the symptom] is in check, because going to the hospital and checking the JKM is too much hassle”. The hospital is only 10 minutes walk from her home, yet checking of JKM and digital appointment is too much of a mental burden for her. Such behavior could lead to serious consequences as treatments could be delayed and her health can be negatively impacted by various degrees. 

\subsection{Feeling of Denial by the Society}
Even in cases when seniors do visit the hospital, most of them have to seek assistance when JKM is checked, and in cases when no help is available, they may still give up going to their destination and return home. As described by a participant ``[the receptionist at hospital] said no JKM no access, so I might as well not go...in the end one doctor stepped in and helped me [display JKM]''. This participant described the recipient as ``evil'', and this feeling of opposition is not uncommon. Many seniors expressed feelings of being unaccepted and unwanted by the society as they are unable to follow the rules of the society and further, not even considered when these rules were made in the first place.

``[some younger people] consider us as burdens to the society, that they have to pay tax to feed us. They hope us better die. They don’t know that one day they will age too.''

\subsection{Increased Dependencies on Others }
As JKM increases seniors’ level of dependency on others, it also consumes their level of confidence. Some participants ask their children to find and screenshot their JKM for them. Sometimes a recent screenshot of JKM is accepted, so the participants will have to ask their children to do this for them often. Sometimes a screenshot is not accepted, then the participants would have to turn to the staff for help. For these participants, a feeling of insecurity is always around them, for they do not know when their screenshot will be rejected. For participants who want to learn how to find JKM, it is not much easier, either. “[my child] showed me three times how to find [JKM] but I still can not remember how to do it. Asking [my child] one or two times is fine; after three or four times they must be annoyed’’ said a participant. This concern about the relationship with children and self-awareness is not uncommon. Not being able to use an unfamiliar technology which is officially widely required for travel does hurt some seniors’ level of confidence.

\section{Senior-unfriendly Designs of JKM}
Recognizing that the JKM created an extract barrier for seniors' travel intentions, we now seek to understand the set of design flaws that of the current JKM. We first analysis the relationship between respondents' personal assessments of their familiarity of smartphones and its relations to their perceived usability of JKM's UI. We then zoom into several common obstacles that prevents the elder individuals from using the JKM in a convenient way.

\subsection{Smartphones Knowledge and JKM Satisfactory}
\begin{figure}[h]
    \centering
    \includegraphics[width=0.45\textwidth]{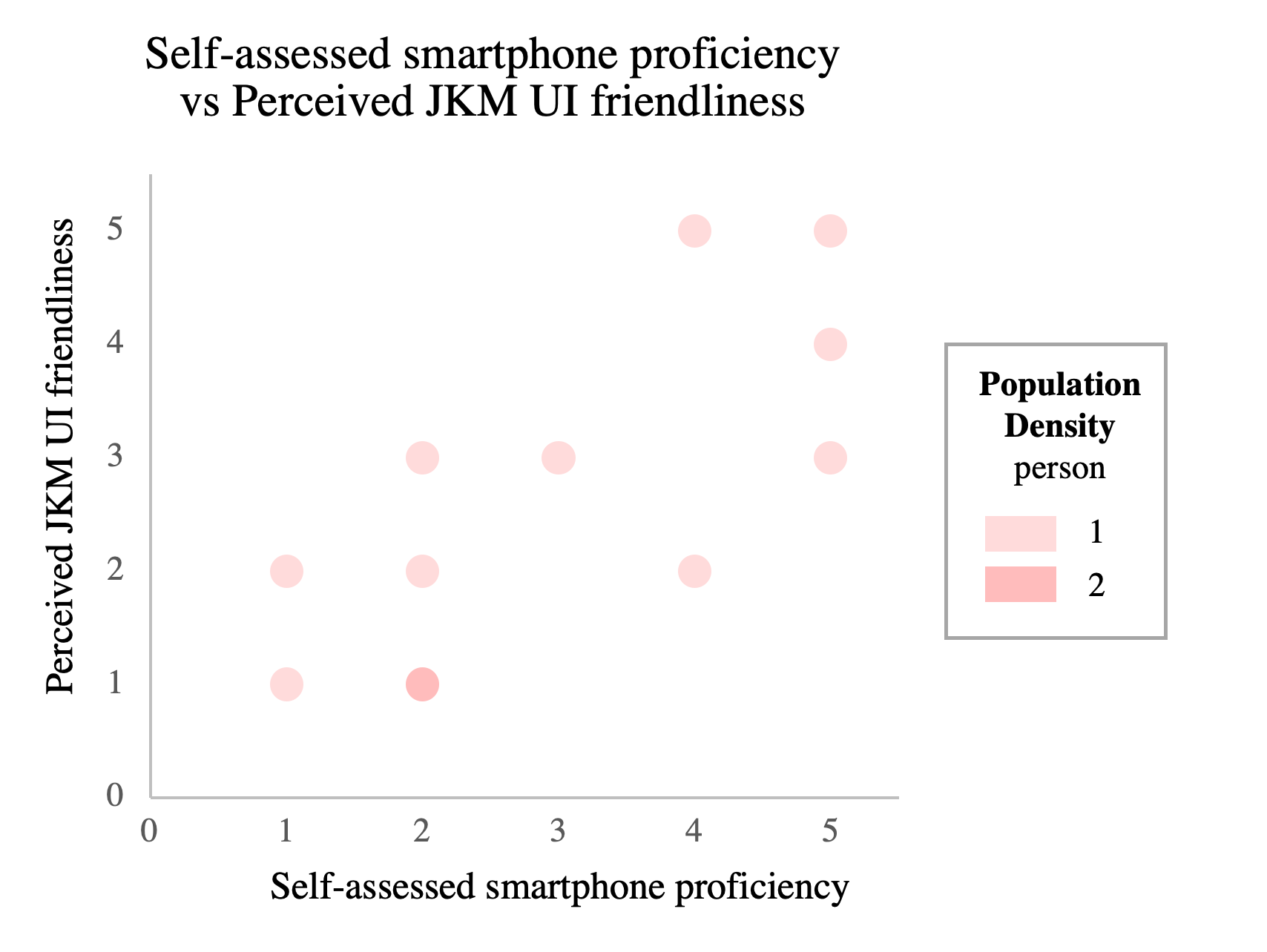}
    \caption{The respondents' self-assessed smartphone proficiency (x axis) and their perceived satisfactory on JKM's UI (y axis)}
    \label{plot:fvui}
\end{figure}
We first want to figure out whether the frustrations about the JKM UI is only restricted within the group of senior citizens who are have little knowledge about smartphones in general. In Fig.~\ref{plot:fvui}, we plotted the self-assessed smartphone knowledge in the x axis along with UI satisfactory (y axis). Here we do see a correlation betwen the general satisfactory and the skillfulness of the users. However, in many situation even participants who are tech savvy find JKM hard to use. Fig.~\ref{plot:fvui} shows that about 80$\%$ of participants with various familiarity with smartphones perceive the UI of JKM as very hard to use. 
  
\subsection{Commonly Used Apps Among Respondents}
\begin{figure}[h]
    \centering
    \includegraphics[width=0.45\textwidth]{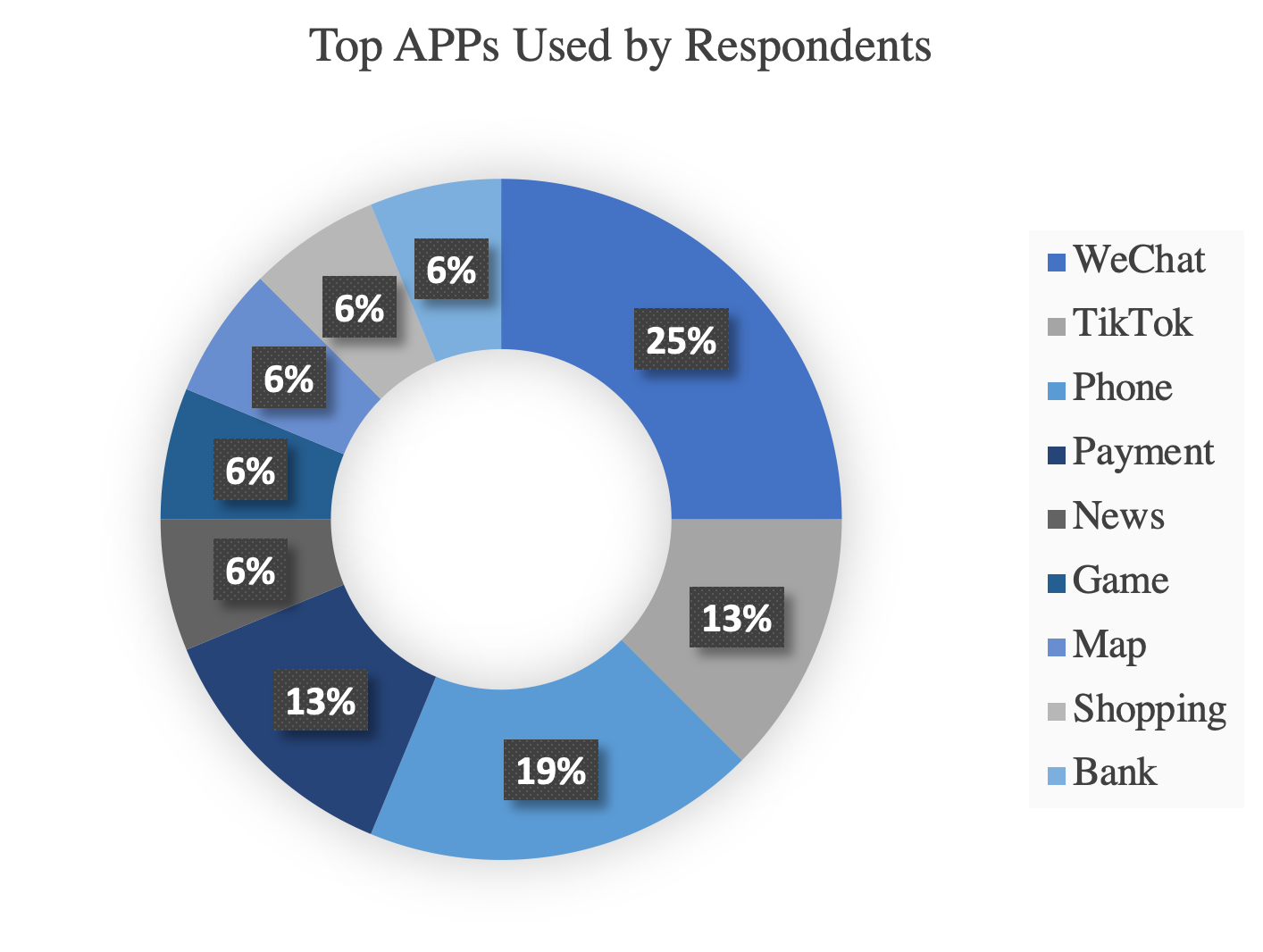}
    \caption{Top APPs Used by Respondents}
    \label{chart:ca}
\end{figure}
We also analyzed the most commonly used apps among the respondents. The most commonly used APPs among the participants are WeChat (50$\%$, social media), Phone (37.5$\%$, essential phone call app), TikTok (25$\%$, short video) and Ali Pay (25$\%$, mobile payment). As JKM is embedded into WeChat and Alipay, this does means that the majority of the users ($75\%$) are already familiar with the entry points of the main app that hosts the JKM. However, even that's the case respondents still find them difficult to use, probably because the usecase of the JKM deviates from the main day-to-day applications of these apps. And even seniors who have already had experiences with those apps still need to learn how to operate JKJ with those platforms. 

\subsection{Accessibility and Availability of JKM}
Other aspects that make using JKM challenging for seniors include complex entrance, inaccessibility without internet, and frequent updates. JKM’s entrance is within WeChat or AliPay, whose main function is social networking and mobile payment. JKM is listed as an icon, along with many other features, within WeChat mini programs/front page of AliPay (screenshots). This makes it hard for the participants to find the JKM and contributes to their learning curve - ``[my child] showed me three times how to find [JKM] but I still do not remember how to do it'' said a participant. Several participants also expressed frustrations about not being able to load JKM when there is low cellular signal. They have had experiences when they were blocked at the entrance of their destination not because they did not know how to show their JKM, but because of poor internet. Another commonly mentioned frustration is the frequency of the updates. JKM expires in several days and when it does, one has to manually update it by sending a text to their phone containing a verification code and then entering it in JKM. Users usually will not know when this update needs to take place until when they need to present their JKM, so this lengthy process can slow down their travel unexpectedly. In addition, a participant expressed concern in not being able to enter the verification code in time which is only valid for 3 minutes. 

\section{Improvement Ideas for JKM}
After we understand why the current design of JKM makes seniors frustrated, we now seek to explore ideas to improve it. We first analyze the improvement ideas based on our survey data. We will now show how these improvements can really have an impacts on seniors' willingness to travel.

\subsection{Most Requested Features}
\begin{figure}[h]
    \centering
    \includegraphics[width=0.45\textwidth]{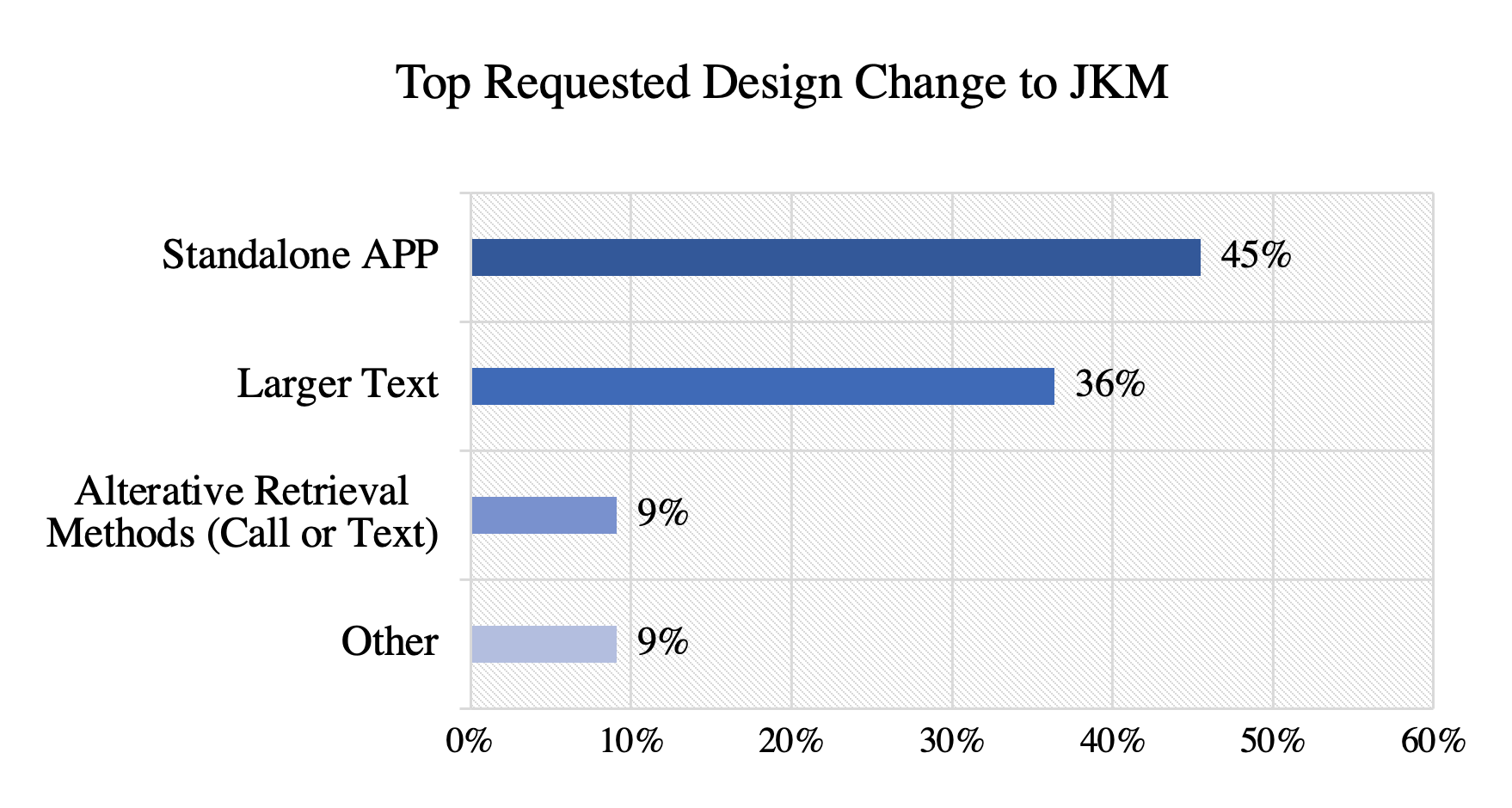}
    \caption{Top requested features by our senior respondents}
    \label{chart:pij}
\end{figure}

\begin{figure*}[t]
\begin{multicols}{3}
    \includegraphics[width=.33\textwidth]{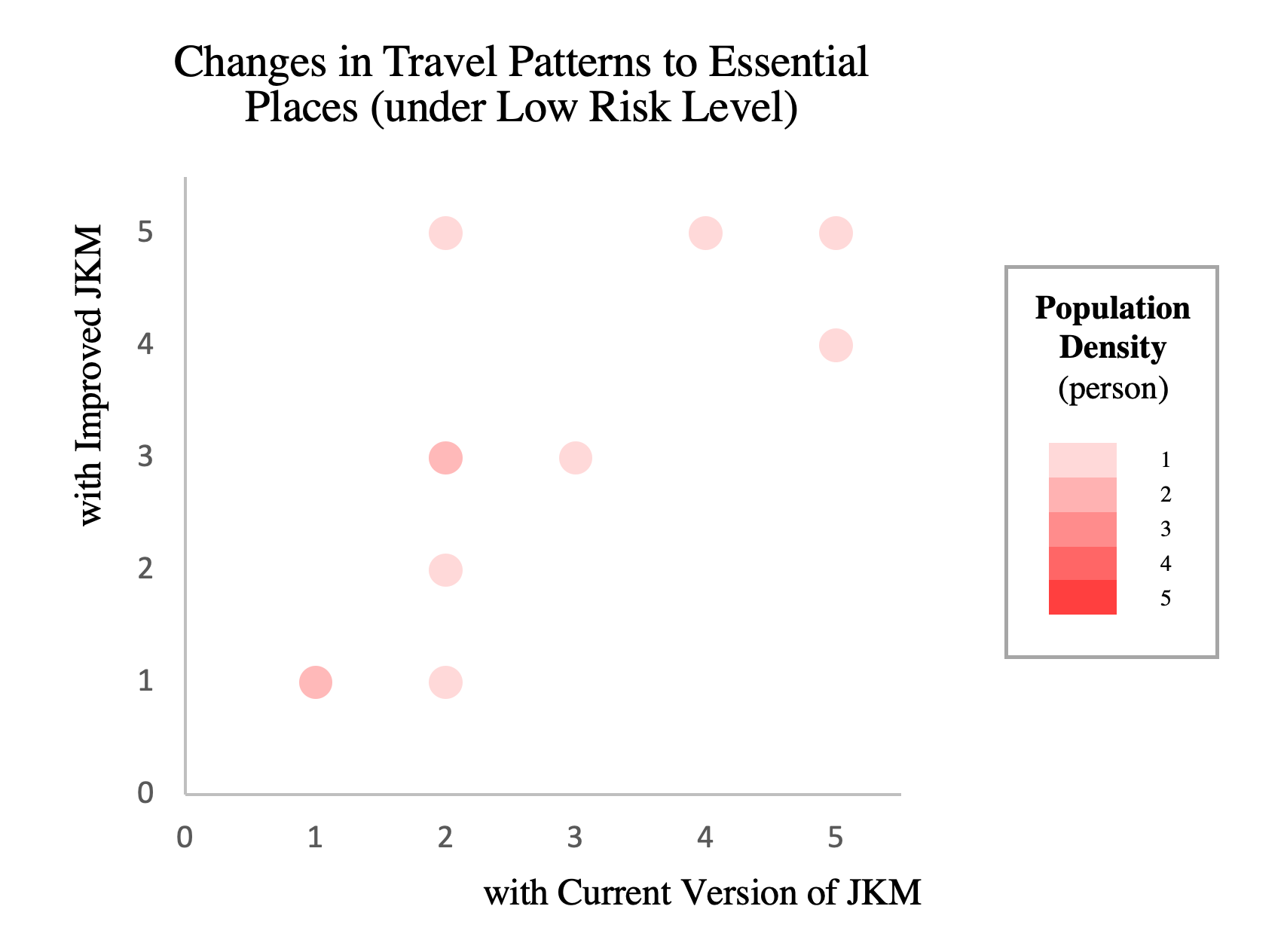}\par
    \includegraphics[width=.33\textwidth]{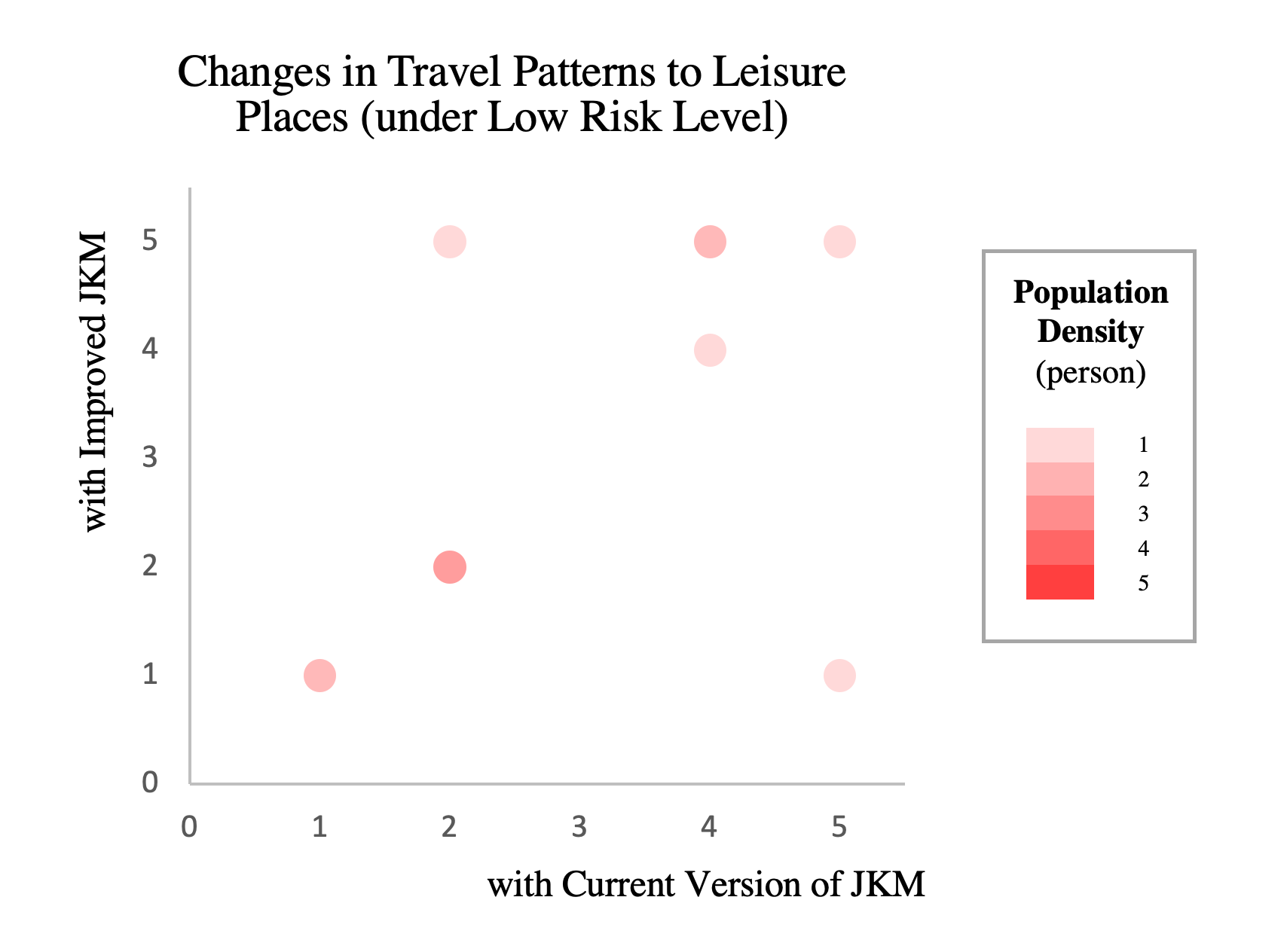}\par
    \includegraphics[width=.33\textwidth]{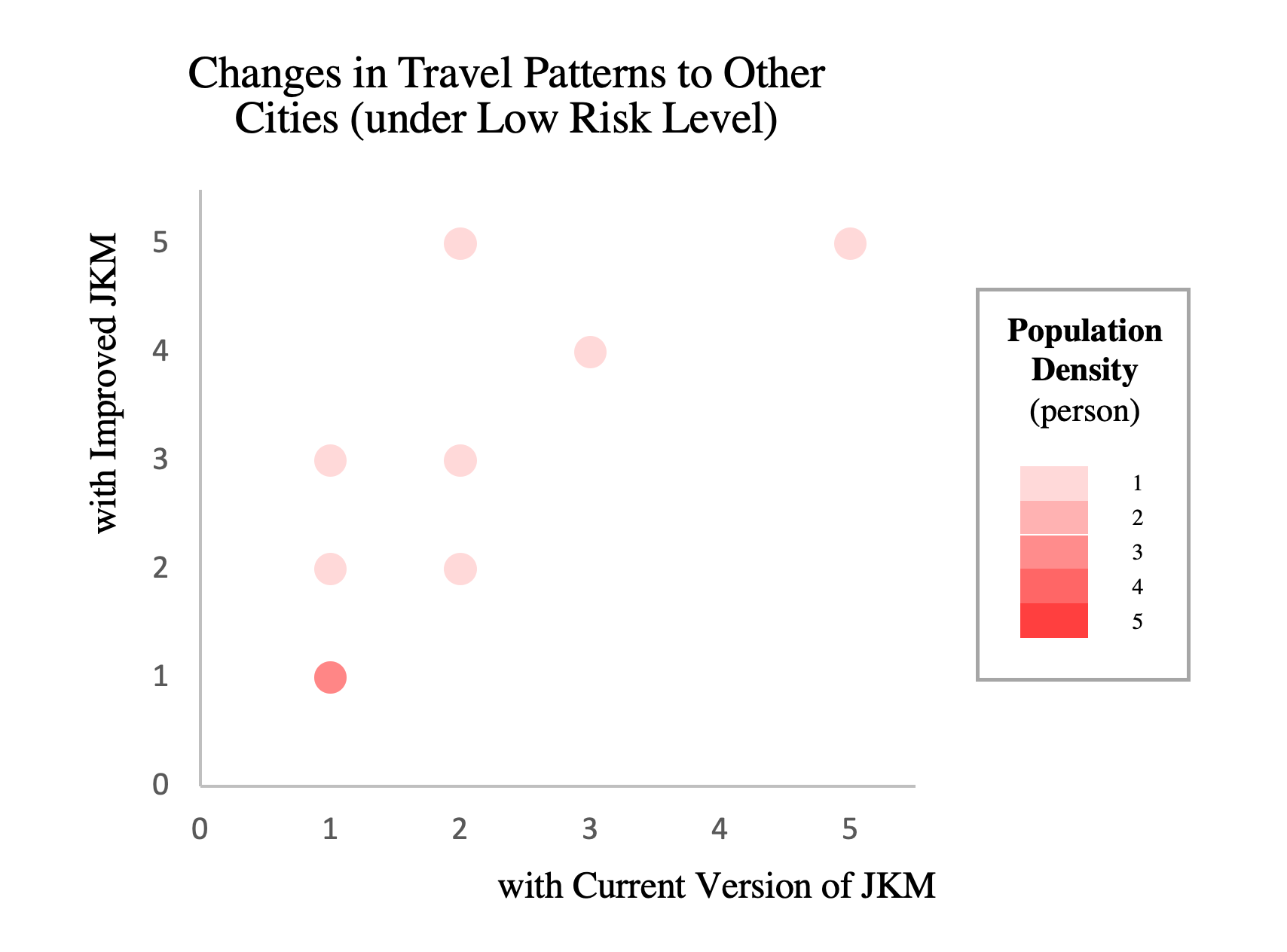}\par
\end{multicols}
\caption{Changes in Travel Patterns with Current Version of JKM vs Improved JKM, Under Low Risk Level, to a) Essential Places, b) Leisure Places, and c) Other Cities}
\label{chart:bvso}
\end{figure*}

We analyzed the top requested features based on our survey data. Fig.~\ref{chart:pij} shows a summary of the results identified by the survey participants. 

\subsubsection{JKM as A Standalone APP}
Many participants mentioned that seniors would like JKM to be as simple as possible. Therefore, the number one requested feature is to make JKM an individual APP, minimizing the steps it takes to find the JKM. This is opposed to the current procedure of retrieving it from Wechat or Alipay, which are two apps that are designed for other purposes. Especially for senior citizens who have presbyopia, making it as a standalone app will eliminate the troubles of finding it within the other platforms.

\subsubsection{Receives JKM Status via Text or Call}
Anther top requested feature from senior citizens is to retrieve JKM using an more traditional approach such as sms or phone call. This method elaborates on seniors’ existing knowledge on the phone. Cell phones are produced decades before smart phones are, and telephones have an even longer history. As a result, a large number of seniors are familiar with texting and most of them are able to make phone calls. Therefore, it would be easy for seniors to be able to receive their JKM status in an alternative way, for example to text or call a certain number. This would benefit people with visual impairment as well.

\subsubsection{Larger Text on the UI}
36$\%$ of the participants mentioned larger text on the UI as an improvement, indicating the current texts are not big enough for seniors. As vision declines over years, larger text can help seniors use JKM more easily.

\subsection{Projected Improvements on Travel Patterns}

As we make improvements on the interactive design of the JKM, we seek to estimate its potential impacts on seniors' travel patterns by asking the same question by with the hypothetical setting of UI being improved. Fig.~\ref{chart:bvso} shows that if JKM is better designed, participants’ desire to travel to essential places, leisure places and other cities all increase, when all else are equal and COVID risk level is low. 

\subsection{Society Acceptance}
More usable JKM also allows seniors use JKM more easily which helps improve their confidence level and self sufficiency. Two participants that we interviewed displayed high level of confidence and independence. They live by themselves - ``why would I live with my children? That would be so annoying [to me]'' ``I have two laptops and two phones. I can spend time with them all day long''. Being tech savvy, they are able to amuse themselves with the internet, watch news online and buy stocks on the phone. They feel a close connection to the society. They also do not feel the rush to find a partner to ``take care of them'', like many seniors do in China. ``I need a man who attracts me spiritually, but I have not find him yet.''

\section{Discussions}
There are 5 levels of human needs according to Maslow's Hierarchy of Needs\cite{mcleod2007maslow}. Unfortunately, the imposition of JKM as a prerequisite for entry into most public places is hurting the needs of Chinese senior citizens. Some seniors suppress their physiological needs with less visits to essential places like the hospital. Some seniors have unfulfilled belongingness and love needs as they lose the connection to the society or their very own children. Finally, some seniors feels less esteem and self-actualization as they are unable to understand how to use JKM, nor can they travel to the places by themselves like before without having to depend on others for help.

One important limitation of this study is the participation of only seniors in China, who may not be representative of other countries. More case studies in other countries would contribute to the understanding of such inequality in a different social context. Second, users’ self-reported smartphone proficiency may be subjective: in the future, objective tests can be used to evaluate participants’ proficiency. Third, a portion of the interviews were conducted in group format, which may lead to bias such as group think - tendency to agree with others in a group setting. Future interviews can be conducted in private before in groups to avoid group think. Finally, when asking participants about their opinions on a better-designed JKM, we did not show participants a redesigned App with improved user experience. Future works can include prototypes and A/B testing which can help participants visualize the improvement ideas and evaluate them more intuitively. \\

\section{Conclusions}
In this paper, we studied senior citizens’ experience and opinions on JKM, a code which every citizen is required to show upon entry to most public places in China, such as hospital, malls and train stations. We revealed the inequalities induced by technology during the Covid pandemic: access to social infrastructure, social isolation, and self-sufficiency. Reflecting on the feedback from the seniors citizens, we explored possible design improvements to the JKM and the positive impact. 
\bibliographystyle{ACM-Reference-Format}
\bibliography{sample-base}

\appendix

\end{document}